\documentstyle[epsf]{elsart}
\begin{document}

\begin{frontmatter}
\rightline{HEPHY-PUB 694/98}
\rightline{UWThPh-1998-31}
\vskip 1truecm
\title{Confinement and chiral symmetry breaking in heavy-light quark systems}
\author{P. Bicudo$^{a}$, N. Brambilla$^{b}$, E. Ribeiro$^{a}$ and  A. Vairo$^{c}$}
\address{$^{a}$ Departamento de Fisica and CFIF-Edificio Ciencia\\
     Instituto Superior Tecnico, Avenida Rovisco Pais, 1096 Lisboa Codex, Portugal}
\address{$^{b}$ Institut f\"ur Theoretische Physik, Boltzmangasse 5, A-1090 Wien, Austria}
\address{$^{c}$ Institut  f\"ur Hochenergiephysik, \"Oster. Akademie der Wissenschaften\\
     Nikolsdorfergasse 18, A-1050 Wien, Austria}

\begin{abstract}
Assuming a Gaussian approximation for the QCD gluodynamics, all the nonperturbative 
physics can be encoded into two parameters: the gluon correlation length $T_g$ and the gluon condensate $G_2$. 
These parameters are sufficient in order to describe the heavy-heavy quark nonperturbative interaction. 
In this work we adopt the same framework in order to study heavy-light bound states in the non-recoil limit. 
Spontaneous chiral symmetry breaking and a confining chiral non-invariant interaction emerge quite naturally. 
The gap equation is solved and discussed. In particular a relation between the light quark condensate 
and $T_g$ is derived. The energy spectrum for the bound state equation is evaluated and commented. 
\end{abstract}

\vskip 1truecm 
\end{frontmatter}
\newpage        

\pagenumbering{arabic}

\section{INTRODUCTION}
The study of heavy-light quark bound states has absorbed in the last years a huge amount of effort 
either on the experimental side, where a lot of new data have been produced (compare, for instance, 
with \cite{data}), either on the theoretical side, due to the use of heavy quark symmetries \cite{isgurw}. 
Around the limit where the symmetries are exact (infinitely heavy quark) a systematic expansion 
in the small parameters $\Lambda_{\rm QCD}/ M_Q$ and $\alpha_{\rm s}(M_Q)$ hold,  $M_Q$ being 
 the mass of the heavy quark.
This expansion contains the long-distance physics of several observables encoded into few hadronic parameters, 
which in general can be defined in terms of operator matrix elements in heavy quark effective theory (HQET) \cite{hqet}.  
In particular in HQET  the heavy-light  meson mass is given by
$$
M_{q\bar{Q}}= M_{Q}+ \bar{\Lambda} + O \left( {1\over M_Q} \right) \, \,{\rm corrections} .
$$
The parameter $\bar{\Lambda}$ is one of the nonperturbative parameters of the HQET. 
It is the sum of all the contributions coming from the terms independent on the heavy-quark mass $M_Q$. 
$\bar{\Lambda}$ can be fixed on the data. In order to calculate it from QCD one needs some dynamical input. 
This dynamics, being the dynamics  of the light quark and the glue, is inerehently nonperturbative. 
Some approaches resort to calculation via phenomenological potential models \cite{hqetpot,dirpot}, 
sum rules \cite{neubert} or relativistic phenomenological equations \cite{phen}. 
To have a well founded calculation of $\bar{\Lambda}$ is of great importance since its value 
affects the determination of many phenomenological quantities (cf e.g. \cite{bigi}).

An attempt in this direction has been done in \cite{bvdir}. In that work an unified description of the heavy-heavy 
and the heavy-light dynamics in the non-recoil limit was suggested on the basis of the so-called Stochastic 
Vacuum model of QCD \cite{DoSi}. This consists in a Gaussian approximation of the QCD gluodynamics which is assumed 
to be governed by the two-point (non-local) gluon condensate. This assumption is confirmed  by lattice simulations 
data \cite{bbv} in the case of quasi-static quarks. In this way a new dynamical scale explicitly appears: 
the gluon correlation length $T_g$. In \cite{bvdir} the heavy mass limit ($m>T^{-1}_g$)  
was explicitly explored. Under this condition two different situations occur. If $T_g^{-1}$ is bigger 
than all the other scales of the problem, then the one-body limit of the heavy quark potential is recovered 
with the famous Eichten--Feinberg--Gromes spin-orbit scalar-like interaction \cite{feinberg}. 
The other case, $T_g^{-1}$ smaller than all the other scales of the problem, is the typical heavy quark 
sum-rule situation. As a check the well-known expression of the heavy  quark condensate as a function of 
the gluon condensate was obtained. Realistic heavy-light systems are characterized by $m<T^{-1}_g$. 
In this case the quark propagator cannot be considered as a free one and its dynamics is inherentely 
nonperturbative.  For $m=0$ our effective Lagrangian is chiral symmetric.
Then the most appropriate approach is to solve the non-linear bound state equation as suggested 
in \cite{bvdir,simdir} or, in a different language, to determine the physical chiral broken vacuum 
of the problem.  This will be done in the present work by means of the Bogoliubov--Valatin variational 
method. Since the Gaussian approximation of the QCD gluodynamics works very well in the heavy quark 
limit \cite{do}, one expects reasonable results for, at least, B$_{\rm s}$ and D$_{\rm s}$ mesons.
 
More from the  point of view of principles this work would like to shed some light on the QCD nonperturbative 
dynamics. It is clear that when we deal with light quarks, chiral symmetry breaking and confinement 
are the two relevant facts. But it is also true that it is difficult to get hold of both effects. 
On one side, Nambu--Jona--Lasinio models, as well as instantons, explain the main physics of chiral symmetry 
breaking but do not have the color confinement property of QCD. On the other side, the confining models 
that try to explain the light quark phenomenology are mainly of two types: 1) models with a scalar 
confining interaction \cite{phen} that explicitly breaks chiral symmetry but insures the 
spin-orbit interaction  known in the heavy quark limit \cite{feinberg}; 2) models with a vector 
confining interaction \cite{bicudo,ley} which are chiral symmetric and allow spontaneous  breaking 
of chiral symmetry but then are in trouble with the spin-orbit interaction. The heavy-light interaction 
which we propose is obtained cleanly from QCD in a gauge invariant approach and under the assumption 
of the dominance of the bilocal correlator. It has the feature of being able to reproduce the expected 
confining linear potential and the expected nonperturbative spin-orbit coupling in the heavy quark limit 
as well as of being chiral symmetric in the chiral limit \cite{bvdir}. More in detail, our interaction kernel turns 
out to be more complicated than a simple convolution kernel. It depends on two dimensional parameters, 
the gluon condensate and the correlation length that control the nonperturbative physics. The string 
tension $\sigma$ arises as a function of these two parameters in the potential limit.  We stress  
that the presence of these two parameters is relevant in order to reproduce the flux tube structure which 
is the main physical fact of the heavy quark nonperturbative dynamics. For heavy-light systems chiral symmetry 
and confinement appear to be strongly related as also suggested by lattice simulations at non-zero temperature 
where confinement and chiral symmetry breaking show up  at the same temperature \cite{latchi}. This is because 
in the proposed approach chiral symmetry breaking effects emerge in the physical framework of a quark-antiquark 
bound state.  As a consequence chiral symmetry breaking  modifies the binding interaction, which 
 turns out to be non--chiral invariant on the physical vacuum. This is very different with respect to the traditional 
Dyson--Schwinger methods where chiral symmetry breaking is associated with the (gauge dependent) non-linear 
dynamics of a  single quark \cite{roberts}. But it  differs also from the 
traditional phenomenological motivated Dirac equations used in the literature for describing 
heavy-light systems in the non-recoil limit \cite{phen}. In those works a scalar confining interaction 
which explicitly breaks chiral symmetry is introduced by hand and phenomenologically justified. 
Here a chiral non-invariant binding interaction emerges naturally on the chiral broken 
physical vacuum. The main pieces of the puzzle seem to go at the right place.

In the present work a simplified expression for the non-local gluon condensate with respect to the parameterization 
derived from lattice simulations \cite{lat} will be used \cite{dm}.  This is only a technical assumption which 
in this exploratory work helps to increase the transparency of the calculation, leaving the main physics untouched. 
The main point is that we are able to give an unified description of the heavy-heavy and of the non-recoil heavy-light 
bound states in terms of few parameters (the gluon condensate, the gluon correlation length) determined 
by the nonperturbative dynamics of QCD. 

The plan of the paper is the following. In section 2 we set up the formalism and discuss with some extension 
the Gaussian approximation of the QCD gluodynamics. In section 3 we derive the gap equation and 
evaluate the quark condensate for different values of the mass. In section 4 we write the heavy-light 
bound state equation in the non-recoil limit and calculate the spectrum. Conclusions and outlook are 
given in section 5.

\section{THE HAMILTONIAN}
A system made up by a quark $q$ of mass $m$ and an antiquark $\bar{Q}$ of mass $M_Q$ is described 
by the 4-point Green function:
\begin{equation}
G_{\rm inv}(x,u,y,v) =  \langle 0\vert \bar{q}(y)U(y,v)Q(v)\bar{Q}(u)U(u,x)q(x) \vert 0 \rangle.  
\label{Ginv}
\end{equation}
The Schwinger strings $U(y,x) \equiv {\rm P}\,\exp \left\{ \displaystyle 
ig\int_0^1 ds\, (y-x)^\mu  A_\mu(x + s(y-x)) \right\}$ have been added in order 
to have gauge invariant initial and final bound states. P stands for the path ordering of the color matrices.

\begin{figure}[htb]
\vskip -1truecm
\makebox[2.5truecm]{\phantom b}
\epsfxsize=10truecm
\epsffile{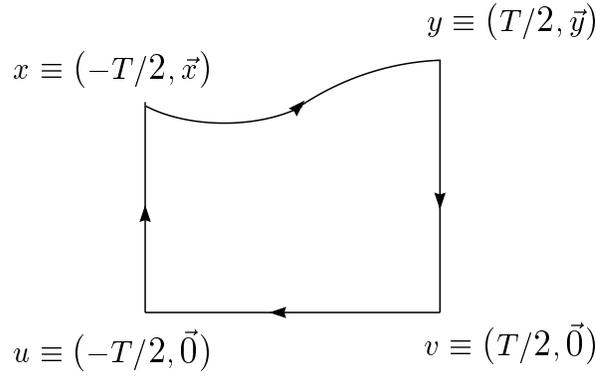}
\vskip -3.5truecm
\caption{The Wilson loop in the static limit of the heavy quark.}
\vskip 0.5truecm
\label{figwh}
\end{figure}

Let us consider the limit in which the antiquark is infinitely heavy ($M_Q \to \infty$). 
The heavy quark behaves then like a static source propagating from $u=(-T/2,{\bf 0})$ 
to $v=(T/2,{\bf 0})$. Let us take  $x=(-T/2,{\bf x})$ and  $y=(T/2,{\bf y})$.
For infinitely heavy  $\bar{Q}$ we have then
\begin{eqnarray}
&~&\!\!\!\! G_{\rm inv}(x,u,y,v) = e^{-iM_Q T} 
\int \!\! {\cal D} q \, {\cal D} \bar{q} \, \Bigg\langle 
\exp\left\{i\int d^4x \,{\cal L}(x)\right\} \bar{q}(y)U(y,v) U(v,u) 
\nonumber\\
&~&~~~~~~~~~~~~~~~~~~~~~~~~~~~~~~~~~~~~~~~~~~ 
\times U(u,x) \left( {1-\gamma^0 \over 2}\right)^{t\,(2)} \!\!\!\! q(x)  \Bigg\rangle,  
\label{Ginv2}\\
&~&\!\!\!\! {\cal L}(x) \equiv  \bar{q}(x) (i\,{D\!\!\!\!/} -m) q(x), 
\nonumber\\
&~&\!\!\!\! \langle {\cal O} \rangle \equiv {1\over {\cal N}} \int {\cal D} A \, e^{i S_{\rm YM}}\,  {\cal O}(A),
\qquad\quad S_{\rm YM} = \hbox{Yang--Mills} ~ SU(3) ~{\rm action}, 
\nonumber
\end{eqnarray} 
where the upperscript $(2)$ refers to the second heavy fermion line,
 $D_\mu = \partial_\mu - i g A_\mu$ and ${\cal N}$ is  a normalization factor. 
A ``natural'' gauge choice is that one which eliminates  the contributions coming from the straight-line 
Schwinger strings $U$ in Eq. (\ref{Ginv2}) \cite{bal85}:
\begin{equation}
A_\mu(x_0,{\bf 0}) = 0, \qquad\qquad x^jA_j(x_0,{\bf x}) = 0.
\label{gauge}
\end{equation}
We then have 
\begin{eqnarray}
\!\!\!\!\!\!\!\!
G_{\rm inv}(x,u,y,v)  &=& e^{-iM_QT}\!\!
\int \!\! {\cal D} q \, {\cal D} \bar{q} \, \left\langle \exp\left\{i\int d^4x \,{\cal L}(x)\right\}\right\rangle 
 \bar{q}(y) \left( {1-\gamma^0 \over 2}\right)^{t\,(2)}  \!\!\!\!q(x).
\nonumber\\
\label{Ginv3}
\end{eqnarray} 
In the Wilson loop language, the Wilson loop $W(\Gamma)$ made up by the heavy quark trajectory, 
a generic light quark path $z$ connecting $x$ with $y$ and the initial and final point Schwinger strings 
(see Fig. \ref{figwh}) simply reduces in the gauge (\ref{gauge}) to the light quark contribution only:
$$
W(\Gamma) \equiv  {\rm Tr \,} {\rm P\,} \exp \left\{ ig \oint_\Gamma dz^\mu A_\mu (z) \right\}   =
{\rm Tr \,} {\rm P\,} \exp \left\{ ig \int_x^y dz^\mu A_\mu (z) \right\}.
$$
Moreover with this gauge choice the gauge field $A_\mu$ can be expressed in terms of the field strength tensor $F_{\mu\nu}$: 
$$
A_0(x) = \int_0^1 d\alpha\, x^k F_{k0}(x_0, \alpha {\bf x}), \quad \quad \quad
A_j(x) = \int_0^1 d\alpha\, 
\alpha \,x^k F_{kj}(x_0, \alpha {\bf x}) . 
$$
We notice that, though this gauge appears to be natural in this problem, its choice is completely arbitrary 
and motivated only by convenience.  Being the starting Green function (\ref{Ginv})
  gauge invariant, by proper handling  
we would obtain exactly the same results within any gauge. 

In order to go on we need an assumption on the gauge field average in Eq.  (\ref{Ginv3}). 
Up to now this is unavoidable since the light quark sector does not provide us with 
an obvious expansion parameter.\footnote{For instance, in the case of heavy quarks the system is  
essentially non-relativistic and we can expand in the quark velocity.} 
We assume that 
\begin{eqnarray}
& & \!\!\!\! \left\langle \exp\left\{ ig\int d^4x \, \bar{q}(x)\gamma^\mu A_\mu(x)  q(x)  \right\} \right\rangle = 
\exp\bigg\{ i\int d^4x \, \bar{q}(x) \gamma^\mu  T^a  q(x)  \langle g A^a_\mu(x) \rangle  
\nonumber\\
& &~~~~~~ - {1\over 2} \int d^4x \int d^4y \,\bar{q}(x) \gamma^\mu  T^a  q(x) 
\bar{q}(y) \gamma^\nu  T^b  q(y)  \langle g A^a_\mu(x) g A^b_\nu(y) \rangle   
\nonumber\\ 
& &~~~~~~ + \hbox{higher order clusters} \bigg\} 
\nonumber\\
& & \!\!\!\! \simeq 
\exp\left\{ - {1\over 2} \int d^4x \int d^4y \, \bar{q}(x) \gamma^\mu  T^a  q(x) 
\bar{q}(y) \gamma^\nu  T^b  q(y)  \langle g A^a_\mu(x) g A^b_\nu(y) \rangle   \right\}.
\label{svm} 
\end{eqnarray}
The first equality is exact. It follows from interpreting the gauge field average as a statistical 
average and from  performing  a cluster expansion \cite{DoSi}. The second is an approximation. 
It corresponds to stop the expansion with the first non-vanishing cluster. 
This approximation was used in this context in \cite{bvdir,simdir}. Although it has been very successful 
in the last years either in applications to the heavy quark potential where it was first 
proposed \cite{DoSi}, as well as in the study of soft high-energy scattering problems (for some recent reviews see \cite{do}). 
Equation (\ref{svm}) is our key assumption. It has been proven to work for quasi-static quarks as the phenomenological 
success of the quoted papers seems to show and as it is confirmed by lattice data (see discussion in \cite{bbv} and \cite{sico}).
For systems with a light quark there is no a priori reason nor to accept it nor to discard it. At this stage  of our understanding 
we will assume that approximation (\ref{svm}) works and see how far we can go with it. In particular in this work 
we will see how chiral symmetry breaking emerges in a heavy-light bound state and how it leads to a non-chiral invariant 
confining interaction. Since we are confident that our assumption works al least for heavy quark bound states we will assume a 
conservative attitude and realistically expect that our picture might be  able to describe B$_{\rm s}$ and D$_{\rm s}$ mesons.  
Nevertheless in the following we will not put a lower bound  on the light quark mass.  

The effective Lagrangian density we will use is therefore 
\begin{eqnarray}
{\cal L}(x) &=&  \bar{q}(x) (i\,{\partial\!\!\!/} -m) q(x)
\nonumber\\ 
&+& {i\over 2} \int d^4y \, \bar{q}(x) \gamma^\mu  T^a  q(x) 
\bar{q}(y) \gamma^\nu  T^b  q(y)  \langle g A^a_\mu(x) g A^b_\nu(y) \rangle, 
\nonumber
\end{eqnarray}
with 
\begin{equation}
\!\!\!\!\!\!\! \langle g A^a_\mu(x) g A^b_\nu(y) \rangle = 
x^k y^l\int_0^1 d\alpha \, \alpha^{n(\mu)} \int_0^1 d\beta\, \beta^{n(\nu)}
\langle g F^a_{k\mu}(x^0,\alpha{\bf x}) g F^b_{l\nu}(y^0,\beta{\bf y})\rangle,  
\label{deff}
\end{equation} 
where $n(0) = 0$ and $n(i) = 1$. The quantities in Eq. (\ref{deff}) are gauge invariant since, due to the gauge choice, 
all the field strength tensors can be thought to be connected by straight-line Schwinger strings.  
The corresponding Hamiltonian is given by:
\begin{eqnarray}
H  &=& \int d^3x \, q^\dagger(x) (-i {\bf \alpha} \cdot {\bf \nabla} +m\beta) q(x)  
\nonumber\\
& & - {i\over 2} \int d^3x \int d^4y \, \bar{q}(x) \gamma^\mu  T^a  q(x) 
\bar{q}(y) \gamma^\nu  T^b  q(y)  \langle g A^a_\mu(x) g A^b_\nu(y) \rangle,  
\label{heff}
\end{eqnarray}
where we have introduced the traditional notation $\beta = \gamma^0$ and $\alpha^i = \gamma^0\gamma^i$. 

We notice that the interaction with the heavy quark source only apparently disappeared from $H$. 
Actually, it manifests itself via the breaking of  translational invariance in the 4-fermion term 
due to the gauge choice (\ref{gauge}). Therefore the effective Hamiltonian given in Eq. (\ref{heff}) 
is simply another way to write the bound state equation already given in \cite{bvdir,simdir}. 
It can, and in principle it should, be used as it is. The whole nonperturbative 
physics is contained into the non-local gluon condensate $\langle g^2 F^a_{\mu\nu}(x) F^b_{\rho\lambda}(0)\rangle$ 
and a parameterization of it is known from lattice measurement \cite{lat}. Being interested only in the 
nonperturbative contributions, a good parameterization is (in Euclidean space) 
\begin{equation}
\langle g^2 F^a_{\mu\nu}(x) F^b_{\rho\lambda}(0)\rangle = {1\over 96}\delta^{ab}(\delta_{\mu\rho}\delta_{\nu\lambda}
-\delta_{\mu\lambda}\delta_{\nu\rho} )\langle g^2 F^2(0)\rangle e^{-\vert x \vert/T_g}. 
\label{real}
\end{equation}
In particular it manifests a long range  exponential fall off in $\vert x \vert$ with a gluonic correlation 
length $T_g \sim$ 0.3 $\div$ 0.4 fm. Once the form of the non-local gluon condensate is plugged in Eq. (\ref{deff}), 
the Hamiltonian describing the nonperturbative dynamics of heavy-light mesons in QCD (under the crucial assumption 
(\ref{svm})) is in principle completely defined. 

At this point we are left with the usual technical problems connected with the actual solving of the bound state 
equation. In the following we will bypass all these by not keeping the ``realistic'' lattice parameterization 
of Eq. (\ref{real}), but by making a rough approximation on it.  This will allow us to perform almost 
all the calculations analytically and to take advantage from some existing literature. The approximation will be 
very rough and we do not expect that it will allow us to make quantitative predictions. 
Nevertheless the qualitative features of the Hamiltonian (\ref{heff}) will be preserved. In particular we will see how 
the mechanism of chiral symmetry breaking is expected to work in the bound state framework and how it leads to 
a non chiral invariant effective bound state interaction. Let us, first,  neglect all the perturbative 
contributions to $\langle g^2 F^a_{\mu\nu}(x) F^b_{\rho\lambda}(0)\rangle$. They are irrelevant since they 
do not lead to confinement and to chiral symmetry breaking. We assume that the leading contribution to 
it is given by the electric fields only and that they are allowed to vary only in time. Moreover we approximate 
the exponential fall off in time with correlation length $T_g$ with an instantaneous delta-type interaction. 
This can be done since for light quarks the energy scale $T_g^{-1}$ is expected to be bigger than the other 
scales of the problem. More explicitly our crude approximation is 
\begin{equation}
\langle g A^a_\mu(x) g A^b_\nu(y) \rangle \simeq - i {\delta^{ab}\delta_{\mu 0}\delta_{\nu 0} \over 24}
\,  {\bf x} \cdot {\bf y} \,\delta(x_0-y_0) \, T_g \,\langle g^2 E^2(0)\rangle.
\label{rough}
\end{equation}
A similar assumption can be found in \cite{dm}. The effective Hamiltonian (\ref{heff}) becomes now:
\begin{eqnarray}
H  &=& \int d^3x \, q^\dagger({\bf x}) (-i {\bf \alpha} \cdot {\bf \nabla} +m\beta) q({\bf x})\\
\nonumber  
& & - {1\over 2} \, V^3_0 \int d^3x \int d^3y \, r^2 \,  q^\dagger({\bf x}) T^a  q({\bf x}) 
q^\dagger({\bf y}) T^a  q({\bf y}) 
\nonumber\\
& & + 2 \, V^3_0 \int d^3x \int d^3y \, R^2  \, q^\dagger({\bf x}) T^a  q({\bf x}) 
q^\dagger({\bf y}) T^a  q({\bf y}), 
\label{heff2}\\
V_0^3 &\equiv& -{T_g \over 96}\langle g^2 E^2(0)\rangle =  {T_g \over 96} \pi^2 G_2, 
~~~~~~~~~~~~ G_2 \equiv \langle {\alpha_{\rm s}\over\pi}F^2(0) \rangle, 
\nonumber
\end{eqnarray}
where ${\bf R} = {\bf x}/2 + {\bf y}/2$ and ${\bf r} = {\bf x} - {\bf y}$. Since the interaction is 
instantaneous from now on we will consider the fermion fields as a function of the spatial coordinates only. 
From Eq. (\ref{heff2}) it is very clear the role played by the approximation (\ref{rough}). 
It has allowed us to disentangle trivially in our effective Hamiltonian the self interacting 
part (function of $r$) from the external source interacting term (function of $R$). 
In the starting Hamiltonian (\ref{heff}) with a ``realistic'' lattice  parameterization of the non-local 
gluon condensate, like Eq. (\ref{real}), these terms might be mixed up in a very complicate way.

\section{THE GAP EQUATION}
In this section we concentrate on the chiral properties of the Hamiltonian (\ref{heff}). 
Following \cite{bicudo,ley} we will use a Bogoliubov--Valatin variational method in order to select 
the chiral broken vacuum. Let us expand the quark fields on a trial basis of spinors:  
\begin{equation}
q({\bf x}) = \sum_s \int {d^3k \over (2\pi)^3}  e^{i {\bf k}\cdot {\bf x}}
\left [ u_s({\bf k}) b_s({\bf k}) + v_s({\bf k})d^\dagger_s(-{\bf k})  \right ],
\label{psitri}
\end{equation}
where the trial spinors $u_s$ and $v_s$ satisfy the usual orthonormality relations
$$
u^\dagger_{s'}({\bf k}) u_s({\bf k}) = v^\dagger_{s'}({\bf k}) v_s({\bf k}) = \delta_{ss'}, \qquad\qquad 
u^\dagger_{s'}({\bf k}) v_s({\bf k}) = v^\dagger_{s'}({\bf k}) u_s({\bf k}) = 0.
$$
A possible choice is 
\begin{eqnarray}
u_s({\bf k}) = {1\over \sqrt{2}} \left[ \sqrt{1 + \sin \phi(k)} + \sqrt{1 - \sin \phi(k)} \, {\bf \alpha}\cdot \hat{\bf k} 
\right]u^0_s, \label{us}\\
v_s({\bf k}) = {1\over \sqrt{2}} \left[ \sqrt{1 + \sin \phi(k)} - \sqrt{1 - \sin \phi(k)} \, {\bf \alpha}\cdot \hat{\bf k} 
\right]v^0_s, \label{vs}
\end{eqnarray}
where $u^0_s$ and $v^0_s$ are the usual rest-frame spinors for the free particle on the chiral invariant vacuum. 
In a moving frame the spinors on the chiral invariant vacuum are given by $\sin\phi(k)= m/\sqrt{m^2+k^2}$ 
and $\cos\phi(k)= k/\sqrt{m^2+k^2}$. In particular, in the limiting case  $\phi=0$ the trial spinors 
reduce to the free massless one, while for $\phi = \pi/2$ they reduce to free static sources. 
Let us also define the projectors  
\begin{eqnarray}
\Lambda_+({\bf k}) &\equiv& \sum_s u_s({\bf k})u^\dagger_s({\bf k}) 
= {1\over 2}\left[1 + \beta \sin \phi(k) + {\bf \alpha}\cdot \hat{\bf k} \cos\phi(k)\right] , 
\nonumber\\
\Lambda_-({\bf k}) &\equiv& \sum_s v_s({\bf k})v^\dagger_s({\bf k}) 
= {1\over 2}\left[1 - \beta \sin \phi(k) - {\bf \alpha}\cdot \hat{\bf k} \cos\phi(k)\right]
= 1 - \Lambda_+({\bf k}).
\nonumber
\end{eqnarray}
The operators $b$, $d$ and $b^\dagger$, $d^\dagger$  are annihilation and creation operators 
respectively. They define the trial vacuum. Eventually after minimizing the vacuum energy they will 
define the chiral broken vacuum. Expanding the Hamiltonian (\ref{heff2}) on (\ref{psitri}) we obtain 
\begin{eqnarray}
H &=& {\cal E} + H^r_2 + H^R_2+ H_4 
\label{heff3}\\
{\cal E} &=& {\cal V} \left\{ 3 \int \!\! {d^3k \over (2\pi)^3} {\rm Tr} \left[ 
({\bf \alpha} \cdot {\bf k} + m \beta) \Lambda_-({\bf k}) \right] \right. 
\nonumber\\
& & \left . -  2 \, V_0^3 \int \!\! {d^3k \over (2\pi)^3} \int \!\! {d^3k' \over (2\pi)^3} 
{\rm Tr} \left[  \Lambda_-({\bf k}) \Lambda_+({\bf k'}) \right] 
\int d^3r \, e^{-i({\bf k} - {\bf k'}) \cdot {\bf r} } r^2 \right\}
\label{ee}\\
H^r_2 &=& \int d^3x : q^\dagger({\bf x}) \left( -i {\bf \alpha}\cdot {\bf \nabla} + m \beta \right) 
q({\bf x}):
\nonumber\\
& & -{2\over 3}\, V_0^3 \int d^3R \int d^3r  \, r^2 \int \!\! {d^3k \over (2\pi)^3}  e^{i{\bf k} \cdot {\bf r} } 
\!\! :q^\dagger({\bf x}) \left[ \Lambda_+({\bf k}) - \Lambda_-({\bf k}) \right] q({\bf y}):  
\label{h2r}\\
H^R_2 &=& {8\over 3} \, V_0^3 \int d^3R \, R^2  \int d^3r \int\!\! {d^3k \over (2\pi)^3}   e^{i{\bf k} \cdot {\bf r} } 
\!\! :q^\dagger({\bf x}) \left[ \Lambda_+({\bf k}) - \Lambda_-({\bf k}) \right] q({\bf y}):  
\label{h2R}\\
H_4 &=& - {1\over 2} \, V_0^3 \int d^3R \int d^3r  \, (r^2 - 4 R^2)
:q^\dagger({\bf x}) T^a q({\bf x}) q^\dagger({\bf y}) T^a q({\bf y}): 
\label{h4}
\end{eqnarray}
where $: ~~:$ is the normal ordering operator and $\cal V$ is the volume of the space. 

\begin{figure}[htb]
\vskip -4truecm
\makebox[0.8truecm]{\phantom b}
\epsfxsize=13truecm
\epsffile{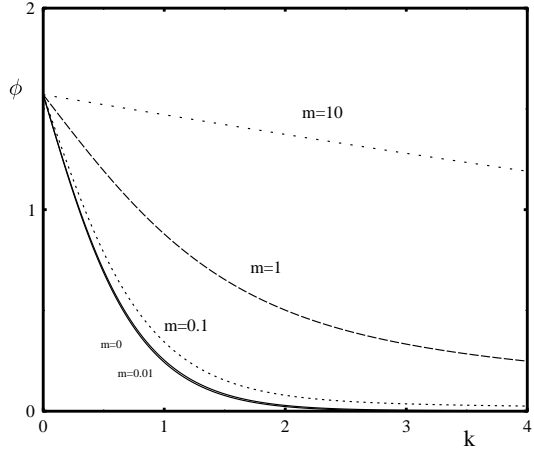}
\vskip -1truecm
\caption{Chiral angles (solution of the gap equation) for different quark masses. 
The momentum $k$ and the mass $m$ are in units of $4 V_0^3 /3$.}
\vskip 1truecm 
\label{figphi}
\end{figure}

$\cal E$ is the vacuum energy. On our trial basis it is a function of the chiral angle $\phi$. 
The Bogoliubov--Valatin variational method consists in choosing $\phi$ which minimizes $\cal E$:
\begin{equation}
\delta {\cal E}(\phi) = 0. 
\label{gap} 
\end{equation}
This is the gap equation. From Eq. (\ref{ee}) more explicitly  we get  
\begin{eqnarray}
\!\!\!\!
m \cos\phi(k) - k\sin\phi(k) + {2\over 3} V_0^3 \left[\phi(k)^{\prime\prime} + {2\over k}\phi(k)^\prime 
+ {2\over k^2} \cos\phi(k) \sin\phi(k) \right] &=& 0.
\nonumber\\
\label{gap2}
\end{eqnarray}
This is just the equation solved in \cite{ley}. As noticed there, it is identical to the Dyson--Schwinger 
equation we would obtain for the self-mass of the light quark interacting via a harmonic 
oscillator potential in the instantaneous ladder approximation. The solution $\phi$, which is a function of $k$, defines 
the new chiral broken vacuum. A plot of $\phi$, solution of Eq. (\ref{gap2}), for different values of the 
mass $m$ is shown in Fig. \ref{figphi}. As the mass increases as much the solution approaches the 
infinitely massive limit $\phi = \pi/2$. Finally, the explicit calculation shows that $\phi$, solution of 
Eq. (\ref{gap2}), not only is a stationary point of the vacuum energy, but also that his vacuum is energetically 
favored with respect to the chiral symmetric one (in the massless case ${\cal E}(\phi) - {\cal E}(0) < 0$). 

The light quark condensate can be calculated explicitly from Eq. (\ref{psitri}):
$$
\langle 0 \vert \bar{q} q \vert 0 \rangle = -{3\over \pi^2} \int_0^\infty dk\, k^2 \sin\phi(k). 
$$
On the solution of Eq. (\ref{gap2}) we get for $m=0$ (see Ref. \cite{ley})  
\begin{equation}
\langle 0 \vert \bar{q} q \vert 0 \rangle = -{1\over 24} \, T_g \, G_2 \times 0.3722 
\label{qbarq}
\end{equation}
The result (\ref{qbarq}) looks appealing. It establishes a connection between the gluon condensate 
and the light quark condensate. The connection is possible since the non-local gluon condensate  
has introduced into the game a finite correlation length $T_g$. Substituting to $T_g$ the lattice 
value 0.35 fm  \cite{lat} and to the gluon condensate $G_2$ the value of 0.048 GeV$^4$ \cite{nar} we get 
$\langle 0 \vert \bar{q} q \vert 0 \rangle \simeq -$ (110 MeV)$^3$ which is a rather low value. 
This is not surprising and is entirely due to our crude assumption (\ref{rough}). If used for static 
sources Eq. (\ref{rough}) would lead to an harmonic oscillator confining potential systematically 
below the expected linear rising potential with slope $\sigma =$ 0.2 GeV$^2$ in the range of interest 
from 0.1 to 1 fm. This suggests that, if we would use the covariant  parameterization of the non-local 
gluon condensate given in Eq. (\ref{real}), that we know from Ref. \cite{DoSi} leads to a phenomenological 
correct linear confinement between static sources, we would presumably enhance the light quark condensate 
to the usually accepted value of $\langle 0 \vert \bar{q} q \vert 0 \rangle \simeq -$ (250 MeV)$^3$. 

Let us  briefly discuss the $H_2^r$ term (\ref{h2r}). Expanding on (\ref{psitri}) we get:
\begin{eqnarray}
H_2^r &=& \sum_{ss'} \int {d^3k \over (2\pi)^3}\left[A(k)\sin\phi(k) + B(k)\cos\phi(k)\right]
\nonumber\\
& & \qquad\qquad\qquad~
\times \left[   b^\dagger_s({\bf k}) b_{s'}({\bf k}) + d^\dagger_{s'}({\bf k}) d_s({\bf k}) \right]\delta_{ss'} 
\nonumber\\
& & \qquad~~~~~~ - \left[A(k)\cos\phi(k) - B(k)\sin\phi(k)\right]
\nonumber\\
& & \qquad\qquad\qquad~
\times \left[   b^\dagger_s({\bf k}) d^\dagger_{s'}(-{\bf k}) + d_{s}(-{\bf k}) b_{s'}({\bf k}) \right]
({\bf \sigma}\cdot\hat{\bf k})_{ss'}, 
\nonumber\\
A(k) &\equiv& m + {2\over 3} \, V_0^3 \,\Delta \sin\phi(k),
\nonumber\\
B(k) &\equiv& k + {2\over 3} \, V_0^3 \left( \Delta \cos\phi(k) - {2\over k^2}\cos\phi(k)\right).
\nonumber
\end{eqnarray}
The stationarity condition (\ref{gap}) or (\ref{gap2}) cancels the so-called Bogoliubov a\-no\-ma\-lo\-us term 
(proportional to $A\cos\phi - B \sin\phi$) which would destabilize the vacuum. On the physical vacuum 
we then have 
\begin{equation}
H_2^r = \sum_{s} \int {d^3k \over (2\pi)^3} E(k) 
\left[   b^\dagger_s({\bf k}) b_{s}({\bf k}) + d^\dagger_{s}({\bf k}) d_s({\bf k}) \right], 
\label{h2r2}
\end{equation}
with $E^2 \equiv A^2+B^2$. Eq. (\ref{h2r2}) simply gives the light quark kinetic energy on the physical vacuum.

\section{THE BOUND STATE EQUATION}
The binding is given in the effective Hamiltonian (\ref{heff3}) by the terms $H_2^R$ and $H_4$. 
As far as the bare $q\bar Q$ mass is concerned we do not need to evaluate $H_4$ matrix elements.
For instance, a term like $\langle q|H_4|q \bar{q}\rangle$  would be responsible for the coupling 
of mesons to the bare $q\bar Q$ state \cite {bicRib2}. Notice that, terms like 
$\langle 0|H^R_2|q \bar{q} \rangle$ which would lead to coupled channels  
and terms contributing only for baryons, have been also neglected. Expanding Eq. (\ref{h2R}) 
on the quark field (\ref{psitri}) and taking the matrix element between  
a one-particle state of momentum ${\bf p}$ and a one-particle state of momentum ${\bf q}$, we have  
\begin{eqnarray}
& &H_2^R({\bf p},{\bf q})_{ss'} \equiv 
\langle 0 \vert b_s({\bf p}) H_2^R  b^\dagger_{s'}({\bf q}) \vert 0 \rangle = 
\nonumber\\
& & \!\!\!\!\! {8\over 3} V_0^3 
u^\dagger_s({\bf p}) \left\{  \beta \sin\left[\phi\!\left({{\bf p} + {\bf q} \over 2}\right)\right]
+ {{\bf\alpha}\cdot\hat{\bf p} +{\bf\alpha}\cdot\hat{\bf q} \over 2} 
\cos\left[\phi\!\left( {{\bf p} + {\bf q} \over 2}\right) \right] \right\}u_{s'}({\bf q})
\nonumber\\
& & \qquad\qquad\qquad \times \left[ -\Delta \, (2\pi)^3\delta^3({\bf p} - {\bf q}) \right] .
\label{h2R2}
\end{eqnarray}
As expected the binding interaction would be chiral invariant ($\sim {\bf\alpha} \cdot\hat{\bf p} 
+ {\bf\alpha}\cdot\hat{\bf q}$) for a massless particle  on the perturbative vacuum ($\phi = 0$). 
While for a infinitely massive particle ($\phi = \pi/2$) chiral invariance would be  maximally broken. 
In our case the solution of the gap equation (\ref{gap2}) gives rise to a binding interaction 
which contains two pieces. One is chiral invariant and the other, proportional to $\beta$, breaks explicitly 
chiral invariance. The existence of such a term is suggested by the spin-orbit structure of the heavy quarkonium  
potential whose relativistic origin may be traced back to a scalar confining Bethe--Salpeter kernel \cite{rep}. 
In a Hamiltonian language this would just correspond to an interaction proportional to $\beta$. 
Indeed such a kind of interaction has been used, also recently, in phenomenological applications 
to the heavy-light spectrum \cite{phen}. As already argued in \cite{bvdir}, what one obtains  
following the approach discussed in this work is an interaction which turns out to be not only 
proportional to $\beta$. It manifests, also under the strong simplifying assumption (\ref{rough}), 
a more complicate structure which interpolates between a chiral invariant 
vector interaction and a scalar interaction generated both by the mechanism of 
chiral condensation and by the initial light quark mass. 

Substituting into Eq. (\ref{h2R2}) the explicit expression for the spinors given by Eq. (\ref{us}) and (\ref{vs}) 
and integrating over a wave function $\Phi$ we get:
\begin{eqnarray}
& & \int\!\!{d^3q\over (2\pi)^3} H_2^R({\bf p},{\bf q}) \Phi({\bf q}) = 
\nonumber\\
& & \qquad {8\over 3} \, V_0^3 \left\{ {1\over 2 p^2} [1-\sin\phi({\bf p})]^2 - {i\over p^2}[1-\sin\phi({\bf p})]
\, {\bf \sigma}\cdot({\bf p}\times{\bf \nabla})  - \Delta \right\} \Phi({\bf p}). 
\nonumber
\end{eqnarray}
Summing up the contributions coming from the pieces $H_2^r$ and $H_2^R$ of the Hamiltonian, the bound state 
equation on the physical vacuum reads
\begin{eqnarray}
& &\left\{ E(p) + {8\over 3} \,V_0^3\left(  {1\over 2 p^2} [1-\sin\phi({\bf p})]^2 
\right.\right. 
\nonumber\\
& & \qquad\qquad\qquad \left.\left. + {2\over p^2}[1-\sin\phi({\bf p})]
\, {\bf S}\cdot{\bf L}  - \Delta \right) \right\}  \Phi({\bf p}) = \bar{\Lambda} \,  \Phi({\bf p}),
\label{bs}
\end{eqnarray}
where we have introduced the spin operator ${\bf S} = {\bf \sigma}/2$ and the orbital
 angular momentum operator 
${\bf L} = {\bf r}\times{\bf p}$. The eigenvalues $\bar{\Lambda}$ of the equation are the energy levels of 
the bound state in the non-recoil limit, i.e. the difference between the mass $M_{q\bar{Q}}$ of the considered 
heavy-light meson and the mass $M_Q$ of the corresponding heavy quark. 

In the heavy quark limit ($m = M_Q\to\infty$, $\phi\to\pi/2$) we get 
$$
\left[ {p^2\over 2 M_Q} - {8\over 3} \, V_0^3 \, \Delta \right]  \Phi({\bf p}) 
= \bar{\Lambda}\,  \Phi({\bf p}). 
$$
This is the bound state equation for two static sources. As already announced,   
assumption (\ref{rough}) leads to a confining potential between infinitely heavy quarks 
which is a harmonic oscillator $\sim V_0^3 r^2$. 

The general spectroscopic properties of Eq. (\ref{bs}) are simple to derive. 
Since $2\,{\bf S}\cdot{\bf L} = J_l^2 - S^2 -L^2$, where ${\bf J}_l$ is the total angular momentum of the 
light quark, and $\Delta = d^2/dp^2 + (2/p)d/dp - L^2/p^2$, the energy levels  are sensitive only 
to $J_l$ and $L$. In particular for each orbital angular momentum quantum number $\ell$ 
we have  two different levels, one with $j_l=\ell+1/2$ (i.e. $2\,{\bf S}\cdot{\bf L} = \ell$) and the other 
with $j_l = \ell-1/2$ (i.e. $2\,{\bf S}\cdot{\bf L} = -1-\ell$). Of course, since in this treatment 
the heavy quark symmetry is exact, there is no dependence of the energy levels on the heavy quark spin.
In Tab. \ref{tabene} we list up to $n=4$ the energy spectrum obtained by solving 
numerically Eq. (\ref{bs}). This spectrum exihibits a small deviation from the pseudo-spin symmetry 
(i.e. the spectrum becomes for J excited states, almost parity independent) which measures 
the extent of chiral symmetry breaking. The value of $\bar{\Lambda}_{\rm u,d} \simeq 0.6$ GeV 
which we obtain with the values of the parameters given in section 3, appears quite reasonable.  

\begin{table}
\makebox[1.5truecm]{\phantom b}
\begin{tabular}{|l|l|l|l|l|l|l|l|}
\hline
$L_l$   & $J_l$ & $J^P$         & n=0   & n=1   & n=2   & n=3   & n=4   \\
\hline
0       & 1/2   & $0^- \ 1^-$   & 2.682 & 6.423 &12.441 &21.002 &32.068 \\
1       & 1/2   & $0^+ \ 1^+$   & 3.524 & 7.927 &14.895 &24.482 &36.634 \\
1       & 3/2   & $1^- \ 2^-$   & 5.254 &10.134 &17.447 &26.606 &39.505 \\
2       & 3/2   & $1^+ \ 2^+$   & 5.431 &10.790 &18.848 &29.533 &42.798 \\
2       & 5/2   & $2^- \ 3^-$   & 7.412 &13.651 &22.364 &33.505 &47.086 \\
3       & 5/2   & $2^+ \ 3^+$   & 7.482 &13.991 &23.204 &35.034 &49.441 \\
3       & 7/2   & $3^- \ 4^-$   & 9.729 &17.303 &27.368 &39.860 &       \\
4       & 7/2   & $3^+ \ 4^+$   & 9.768 &17.511 &27.921 &40.930 &       \\
4       & 9/2   & $4^- \ 5^-$   &12.291 &21.203 &32.591 &46.409 &       \\
5       & 9/2   & $4^+ \ 5^+$   &12.316 &21.343 &32.983 &47.196 &       \\
5       & 11/2  & $5^- \ 6^-$   &15.121 &25.385 &38.086 &       &       \\
6       & 11/2  & $5^+ \ 6^+$   &15.138 &25.488 &38.379 &       &       \\
6       & 13/2  & $6^- \ 7^-$   &18.227 &29.862 &43.876 &       &       \\
\hline
\end{tabular}
\vskip 1 truecm
\caption{ $\bar{\Lambda}$ in $4 V_0^3/3$ units as obtained from Eq. (\ref{bs}) for $m=0$. $J$ and $P$ are respectively 
the total angular momentum and the parity of the meson. }
\vskip 0.5truecm
\label{tabene}
\end{table}

\section{COMMENTS AND CONCLUSIONS}
In this work we have studied the heavy-light non-recoil dynamics of QCD adopting a Gaussian 
approximation for the gluodynamics. All the heavy quark limits ($m>T_g$) were already studied 
in \cite{bvdir} and give the expected results. Here we do not have given any constraint on the light 
quark mass. Chiral symmetry breaking and a chiral non-invariant binding interaction emerge quite 
naturally in our approach and a link is established between chiral symmetry breaking properties 
and confining interaction. In particular with Eq. (\ref{qbarq}) we establish a relation 
between the order parameter of chiral symmetry (the quark condensate $\langle 0 \vert \bar{q} q \vert 0\rangle$) 
and that one which in our framework describes confinement (the gluon correlation length $T_g$). 
A similar relation can be found in \cite{simdir}. 

The actual calculations were performed under the rough approximation (\ref{rough}). 
This is unrealistic since it gives in the heavy quark limit a confining potential 
which is not linear . Moreover all magnetic contributions were not considered. This is expected to affect 
the levels listed in Tab. \ref{tabene}. Nevertheless, as discussed in section 2, we expect 
that the main features presented in this work (chiral symmetry breaking, chiral non-invariant 
binding interaction, relation between the order parameters) will still hold also by using a more 
realistic parameterization of the bilocal gluon condensate, for instance that one suggested 
by lattice simulations. Indeed, the encouraging results obtained in such a simple framework, 
not only support but also make urgent the extension in this sense of the present analysis. 
Work is in progress in this direction.

\section*{Acknowledgments}
We thank H. G. Dosch for many  useful discussions. 
Two of us (N. B. and A. V.) gratefully  thank the members of the Physics Department of the 
Instituto Superior Tecnico of Lisbon for the warm hospitality given to them during the first stage of this work. 
N. B. acknowledges the support of the European Community, Marie Curie fellowship, TMR Contract n. ERBFMBICT961714.

\end{document}